\makeatletter
\declare@file@substitution{revtex4-1.cls}{revtex4-2.cls}
\makeatother
\documentclass{aastex631}
\usepackage{epsfig}
\usepackage{natbib}
\usepackage{amsmath}
\usepackage{amssymb}
\usepackage{enumerate}
\usepackage[toc,page]{appendix}
\usepackage{rotating}
\usepackage{url}
\usepackage{color}

\def\LSI61{LS I+61$^{\circ}$ 303}

\begin{document}

\title{A precessing stellar disk model for superorbital modulations of the gamma-ray binary LS I+61$^{\circ}$ 303}

\author{A. M. Chen}
\affiliation{Tsung-Dao Lee Institute, Shanghai Jiao Tong University, Shanghai 201210, China.}
\affiliation{Key Laboratory of Quark and Lepton Physics (MOE), Central China Normal University, Wuhan 430079, China.}

\author{J. Takata}
\affiliation{Department of Astronomy, School of Physics, Huazhong University of Science and Technology, Wuhan 430074, China.}

\author{Y. W. Yu}
\affiliation{Institute of Astrophysics, Central China Normal University, Wuhan 430079, China.}
\affiliation{Key Laboratory of Quark and Lepton Physics (MOE), Central China Normal University, Wuhan 430079, China.}

\begin{abstract}
Gamma-ray binary LS I+61$^{\circ}$ 303 consists of a neutron star orbiting around a Be star with a period of $P_{\rm orb}\simeq26.5\ {\rm d}$. Apart from orbital modulations, the binary shows long-term flux variations with a superorbital period of $P_{\rm sup}\simeq4.6\ {\rm yrs}$ as seen in nearly all wavelengths. The origin of this superorbital modulation is still not well understood.
Under the pulsar wind-stellar outflow interaction scenario, we propose that the superorbital modulations of LS I+61$^{\circ}$ 303 could be caused by the precession of the Be disk. Assuming X-rays arise from synchrotron radiation of the intrabinary shock, we develop an analytical model to calculate expected flux modulations over the orbital and superorbital phases. The asymmetric two-peak profiles in orbital light curves and sinusoidal-like long-term modulations are reproduced under the precessing disk scenario. The observed orbital phase drifting of X-ray peak and our fitting of long-term X-ray data indicate that the neutron star is likely orbiting around the star with a small eccentricity and periastron phase around $\Phi_{\rm p}\sim0.6$. 
We compare the Corbet diagrams of LS I+61$^{\circ}$ 303 with other Be/X-ray binaries, and the linear correlation in the $P_{\rm sup}-P_{\rm orb}$ diagram suggests that the precession of the Be disk in LS I+61$^{\circ}$ 303 is induced by the tidal torque of its neutron star companion.
\end{abstract}


\section{Introduction} \label{sec:intro}
\LSI61 is a massive $\gamma$-ray binary that contains a compact object (CO) and a rapidly rotating B0Ve star (see \citealt{Marcote2017} for a review of this binary). Due to the unclear nature of CO, \LSI61 had been widely believed to be a microquasar for a long time (e.g.\citealt{Gregory2002b,Bosch-Ramon2004, Gupta2006,Massi2017,Jaron2021}),
although some studies suggest that the CO is an energetic pulsar (\citealt{Maraschi1981,Lipunov1994,Dubus2006,Dhawan2006,Zdziarski2010}).
Recently, \citet{Weng2022} reported detections of transient radio pulses from \LSI61 with a period of $269\ {\rm ms}$, which strongly suggests that the CO is a rotation-powered neutron star (NS)\footnote{However, see \citet{Jaron2024} for a different opinion.}.
It also implies that the broadband emissions from the binary are likely resulting from the interaction between the relativistic pulsar wind with stellar outflows, similar to other $\gamma$-ray binaries like PSR B1259-63/LS 2883 and PSR J2032+4127/MT 91 213
(\citealt{Tavani1997,Takata2012,Takata2017,Petropoulou2018,Chen2019,Chen2022}).

Since its discovery by \citet{Gregory1978}, \LSI61 has been intensively monitored by many telescopes across the whole electromagnetic spectrum
(e.g., \citealt{Hutchings1981,Taylor1982,Taylor1984,Paredes1997,Paredes2007,Albert2006,Albert2008,Chernyakova2006,Chernyakova2012,Chernyakova2023,Acciari2008,Acciari2009,Acciari2011,Hadasch2012,Aleksic2012,Jaron2014,Archambault2016,Marcote2016,Zhang2024}).
The system shows orbital modulated radiation from radio to very-high-energy (VHE) $\gamma$-rays with a period of $P_{\rm orb}= 26.496\ {\rm d}$.
\citet{Anderhub2009} conducted a simultaneous multiwavelength observation of the binary, and they found a significant flux correlation between X-rays and VHE $\gamma$-rays. Orbital light curves in these two bands show a prominent peak around phase 0.6-0.7 and a second broader peak around phase 0.8-1.1. \citet{Chen2022} proposed that this double-peak profile is due to interactions of the pulsar wind with the inclined Be disk.



In addition to orbital modulations, \LSI61 also shows superorbital variations with a period of $P_{\rm sup}\simeq 4.6\ {\rm yr}$ as seen at nearly all wavelengths, including radio (\citealt{Gregory2002a,Marcote2016,Jaron2024}),  optical (\citealt{Paredes-Fortuny2015}), X-ray (\citealt{Li2011,Li2012,Chernyakova2017,DAi2016}),  GeV $\gamma$-ray (\citealt{Hadasch2012,Ackermann2013,Xing2017,Chernyakova2023}), and possibly TeV $\gamma$-ray (\citealt{Ahnen2016}).
The long-term multiwavelength modulation properties have been investigated in great detail by \citet{Chernyakova2012}, \citet{Saha2016}, and \citet{Jaron2021}.  In particular, \citet{Chernyakova2012} found that the X-ray peak shows a systematic drift from the orbital phase of 0.35 to 0.75 within one superorbital period. They also discovered a constant time delay between the X-ray and radio peaks, indicating different emitting regions in these two bands. A detailed study of systematic phase-offsets in radio, X-ray, GeV, and TeV $\gamma$-rays are also presented in \citet{Jaron2021}.
\citet{Saha2016} found that long-term flux variations as seen by \textit{Fermi}/LAT, \textit{RXTE}/PCA, \textit{Swift}/XRT, and Owens Valley Radio Observatory can be well fitted by a sinusoidal function of the superorbital period, with peak positions and amplitudes varying significantly with the binary phase. 

The origin of superorbital modulations in \LSI61 is far from well understood.
\citet{Zamanov1999} suggested that long-term modulations could be due to cyclical changes in the Be star's mass-loss rate or disk density. Alternatively, under the microquasar scenario, \citet{Massi2014} proposed that the superorbital period is caused by the beat frequency between the orbital motion of the CO with $P_{\rm orb}=26.496\ {\rm d}$ and the precession of the microquasar jet with a period of $P_{\rm jet}=26.926\ {\rm d}$. However, this scenario may be incompatible with the detection of radio pulsations and the lack of accretion signals in X-rays (e.g., \citealt{Sidoli2006,Chen2022,Saavedra2023}).

Superorbital periods are seen in many X-ray binaries (XRBs; e.g., \citealt{Sood2007,Kotze2012,Townsend2020}) but are rare in $\gamma$-ray binaries. \LSI61 is the only $\gamma$-ray binary with long-term modulations discovered so far. For Be/XRBs, it has been suggested that the optical superorbital variability is related to the formation and depletion of the Be disk (\citealt{Alcock2001,Suffak2022}) or the precession and warping of the disk due to the tidal effect of the CO companion (\citealt{Rajoelimanana2011}). Those processes can modulate accretion rates onto the CO, leading to long-term modulations of X-rays. Alternatively, \citet{Martin2023} proposed that the stellar spin-axis precession can also drive superorbital modulations in Be/XRBs, particularly for those with shorter orbital periods.

\LSI61 shares similar properties as those Be/XRBs, although the detection of radio pulsation indicates its broadband emission is more likely to be powered by the rotational pulsar instead of the accretion process.
In this paper, we will investigate the disk precession scenario for the superorbital modulations of \LSI61 under the interaction model between the pulsar wind and stellar outflows.
We only focus on the X-ray emission since it is widely believed to be produced by synchrotron radiation in the intrabinary shock (IBS), while emission processes in other wavebands (radio and $\gamma$-rays) are much more complicated and still underdebated.
The paper is organized as follows.
Section 2 describes the disk precession geometry and the emission model for LS I+61$^{\circ}$ 303. 
In Section 3, we present expected X-ray flux modulations over the orbital and superorbital periods and compare observational data. 
We plot the Corbet diagrams of \LSI61 with other Be/XRBs and discuss possible physical origins that drive the precession of the disk in Section 4.
Finally, a brief summary of our emission model is provided in Section 5, and we also discuss the observational properties in other wavebands.

\section{Model description}

\subsection{Precessing disk}

The stellar outflows of Be stars are characterized by two components: a fast-moving dilute polar wind and a slow-moving dense equatorial disk (\citealt{Waters1986}). The exact density and velocity distribution profiles of stellar outflows can be very complicated, particularly for the decretion disk (e.g. \citealt{Torres2012}). Here, we adopt a simplified outflow model for Be stars used in \citet{Petropoulou2018} and \citet{Chen2019}. Assuming that stellar outflows are axisymmetric along the disk normal, we model the ram pressure of stellar outflows as:
\begin{eqnarray}\label{pw}
  p_{\rm w}(R,\theta)&=&p_0R^{-2}(1+G\mid\cos\theta\mid^m),
\end{eqnarray}
where $R$ is the radial distance from the star, $\theta$ is the latitude angle measured from the stellar equator, and $p_0$ is the polar wind ram pressure.
The disk is described by the equator-to-pole pressure contrast $G$ and the confinement index $m$. A higher value of $m$ means a well-confined disk (\citealt{Chen2022}).

\begin{figure}
  \centering
  \includegraphics[width=0.9\textwidth]{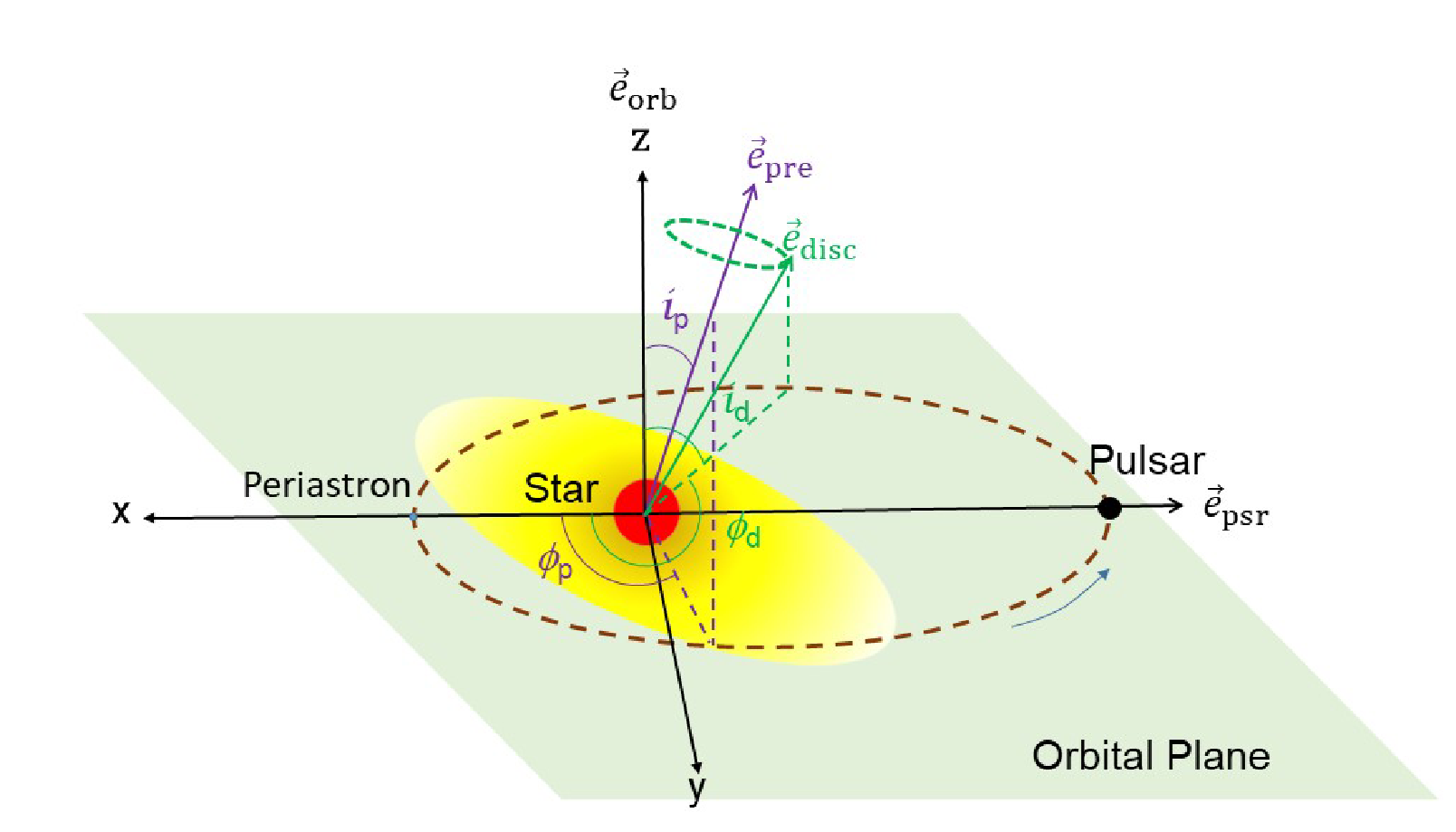}
  \caption{Schematic diagram of the stellar disk precession scenario (not to scale). The normal vector of the stellar disk (solid green arrow) rotates along the precession axis (solid purple arrow) with the trajectory indicated by the dashed green circle. } \label{fig:geometry}
\end{figure}

We attribute superorbital modulations of \LSI61 to the interactions of pulsar wind with a precessing Be disk.
As illustrated in Figure. \ref{fig:geometry}, we define a Cartesian coordinate system with the origin on the stellar center, $x$-axis toward the periastron, and $z$-axis along the orbital axis. Then, the unit vectors of the pulsar and the disk precession axis are (e.g., \citealt{Chen2021a}):
\begin{eqnarray}\label{vecs}
\nonumber \boldsymbol{e}_{\mathrm{psr}}
&=&\mathbf{R}\left[ \boldsymbol{e}_z,\phi \right] \boldsymbol{e}_x\\
&=&\left( \cos \phi ,\sin \phi ,0 \right),
\\
\nonumber \boldsymbol{e}_{\mathrm{pre}}
&=&\mathbf{R}\left[ \boldsymbol{e}_z,\phi _{\mathrm{p}} \right] \mathbf{R}\left[ \boldsymbol{e}_y,-\left( \pi /2-i_{\mathrm{p}} \right) \right] \boldsymbol{e}_x\\
&=&\left(\sin i_{\rm{p}}\cos \phi_{\rm{p}}, \sin i_{\rm{p}}\sin \phi_{\rm{p}}, \cos i_{\rm{p}} \right),
\end{eqnarray}
where $\mathbf{R}[\boldsymbol{e}_i, \xi]$ is the rotation matrix around the unit vector $\boldsymbol{e}_i$ with the counterclockwise rotated angle of $\xi$, $\phi$ is the NS's true anomaly, $i_{\rm p}$ and $\phi_{\rm p}$ are the inclination angle and the true anomaly of the precession axis, respectively.
The normal vector of the precessing disk at a given time $t$ is
\begin{eqnarray}\label{disc}
\boldsymbol{e}_{\mathrm{disk}}\left( t \right) &=& \mathbf{R}\left[ \boldsymbol{e}_{\mathrm{pre}},\dot{\omega} (t-t_0) \right] \boldsymbol{e}_{\mathrm{disk},0},
\end{eqnarray}
with
\begin{eqnarray}
\nonumber \boldsymbol{e}_{\mathrm{disk},0}
&=&\mathbf{R}\left[ \boldsymbol{e}_z,\phi _{\mathrm{d},0} \right] \mathbf{R}\left[ \boldsymbol{e}_y,-\left( \pi /2-i_{\mathrm{d},0} \right) \right] \boldsymbol{e}_x \\
&=&\left(\sin i_{\rm{d,0}}\cos \phi_{\rm{d,0}}, \sin i_{\rm{d,0}}\sin \phi_{\rm{d,0}}, \cos i_{\rm{d,0}} \right),
\end{eqnarray}
being the direction of the disk normal at the reference time $t_0$, $\dot{\omega}=2\pi/P_{\rm pre}$, and $P_{\rm pre}$ is the precession period.

\subsection{Intrabinary shock}

In $\gamma$-ray binaries, the relativistic pulsar wind will be terminated by the intense outflows from the massive star, forming a pair of shocks. The contact discontinuity of shocks is determined by the dynamical balance of two winds:
\begin{eqnarray}\label{dyn}
  \frac{L_{\rm sd}}{4\pi r_{\rm s}^2c}&=&p_0R_{\rm s}^{-2} (1+G\mid\cos\theta\mid^m),
\end{eqnarray}
with $L_{\rm sd}$ being the NS's spin-down power, $c$ is the light speed, and the latitude angle is given by:
\begin{eqnarray}\label{theta}
  \theta(t) &=&\pi /2-\arccos \left[ \boldsymbol{e}_{\text{disk}}(t) \cdot \boldsymbol{e}_{\text{psr}}(\phi) \right].
\end{eqnarray}
In the above equation, we need to express the true anomaly $\phi$ as a function of time $t$ (\citealt{Battin1999}):
\begin{eqnarray}
    \phi(t) &=& M+2\sum_{k=1}^{\infty}{\frac{1}{k}\left[ \sum_{n=-\infty}^{\infty}{J_n\left( -ke \right) \beta_e ^{\left| k+n \right|}} \right] \sin kM},\\
    M(t)&=&M_0+\frac{2\pi}{P_{\mathrm{orb}}}\left( t-t_0 \right),
\end{eqnarray}
where $\beta_e = e/(1+\sqrt{1-e^2})$, $e$ is the orbital eccentricity, $J_{n}$ are Bessel functions of the first kind, and $M_0$ is the mean anomaly at the epoch of $t_0$.

It is useful to introduce the momentum ratio of two winds as (\citealt{Chen2019}):
\begin{eqnarray}\label{eta}
  \eta(t)&=&\frac{L_{\rm sd}/c}{4\pi p_0(1+G\mid\cos\theta(t)\mid^m)},
\end{eqnarray}
therefore, the distances of the IBS apex from the NS and star can be simply written as:
\begin{eqnarray}
  r_{\rm s}(t)&=&d(\phi) \frac{\eta(t)^{1/2}}{1+\eta(t)^{1/2}},
\end{eqnarray}
and
\begin{eqnarray}
  R_{\rm s}(t)&=&d(\phi) \frac{1}{1+\eta(t)^{1/2}},
\end{eqnarray}
respectively. The binary separation is
\begin{eqnarray}
  d(\phi) &=& a\frac{1-e^2}{1+e\cos{\phi}},
\end{eqnarray}
with $a$ being the semi-major axis.

\subsection{Synchrotron radiation}
The IBS will compress the pulsar wind magnetic field in the downstream region and accelerate the charged particles (which are mainly electron-positron pairs, hereafter $e^{\pm}$ pairs) to VHE energies. The accelerated $e^{\pm}$ pairs move in the magnetic field and emit synchrotron photons.

The magnetic field strength in the downstream of IBS is:
\begin{eqnarray}\label{B}
  B(t)=\left(\frac{2\epsilon_B L_{\rm sd}}{r_{\rm s}(t)^2c}\right)^{1/2},
\end{eqnarray}
where the magnetic equipartition parameter $\epsilon_B$ is related to the pulsar wind magnetization factor $\sigma$ as $\epsilon_B\simeq9\sigma/2$ for $\sigma<0.1$ or $\epsilon_B\simeq1/2$ for $\sigma>0.1$ (\citealt{Kennel1984}). It is obvious that a smaller value of $r_{\rm s}$ during the disk passages will lead to an increase in the magnetic field of IBS.

The $e^{\pm}$ pairs from pulsar wind are expected to be accelerated into a power-law distribution. We write the pair injection rate as (\citealt{Kirk1999}):
\begin{eqnarray}
  Q\left( \gamma \right) &\simeq& \frac{\epsilon _{\rm e}L_{\mathrm{sd}}\left( p-2 \right)}{\gamma _{\min}^{2-p}m_{\rm e}c^2}\cdot \gamma ^{-p},\quad \gamma\geq\gamma_{\min},
\end{eqnarray}
where $\epsilon_{\rm e}$ is the fraction of the pulsar luminosity being transferred to accelerated $e^{\pm}$ pairs, and $m_{\rm e}$ is electron mass. Assuming that the unshocked pulsar wind contains mono-energetic $e^{\pm}$ pairs with Lorentz factor of $\gamma_{\rm w}$, the minimum Lorentz factor of accelerated $e^{\pm}$ pairs is $\gamma_{\min}=\gamma_{\rm w}(p-2)/(p-1)$.
Since we only consider the expected emission in soft X-ray band (i.e., $0.3-10\ {\rm keV}$), emitting particles considered here are dominated by adiabatic cooling with the timescale of (\citealt{Dubus2006,Kong2012})
\begin{eqnarray}
  \tau_{\rm ad}(t) &\simeq& \frac{3r_{\rm s}(t)}{c},
\end{eqnarray}
and the evolved distribution of cooled $e^{\pm}$ pairs can be simply written as (\citealt{Kirk1999}):
\begin{eqnarray}\label{Ne}
  N(\gamma) &\simeq& Q(\gamma)\tau_{\rm ad}.
\end{eqnarray}

Under the monochromatic approximation, the synchrotron emissivity of a single electron is (\citealt{Petropoulou2018}):
\begin{eqnarray}\label{PX}
  P_{\nu}(\gamma) &\simeq& \frac{4}{3}c\sigma_{\rm T}u_{\rm B}\gamma^2\delta(\nu-\nu_{\rm c}),
\end{eqnarray}
where
\begin{eqnarray}\label{EX}
  h\nu_{\rm c} &\simeq& \frac{3}{2}\frac{B}{B_{\rm cr}}\gamma^2m_{\rm e}c^2,
\end{eqnarray}
is the characteristic synchrotron energy, $u_B=B^2/8\pi$, $B_{\rm cr}=4.414\times10^{13}\ {\rm Gs}$ is the quantum critical field strength, and $\sigma_{\rm T}$ is the Thompson cross section.
The specific synchrotron luminosity is obtained by integrating over the particle distribution:
\begin{eqnarray*}
  L_{\nu} &=& \int{N\left( \gamma \right) P_{\nu}\left( \gamma \right) d\gamma}
\end{eqnarray*}
which gives the analytical expression of synchrotron luminosity in the X-ray frequency range of $[\nu_{\rm X1}, \nu_{\rm X2}]$ as:
\begin{eqnarray}\label{LX}
  L_{\rm X}(t)&\simeq& \mathbb{C} f_{\rm X}\epsilon_{\rm e}\epsilon_{\rm B}^{\frac{1+p}{4}} L_{\mathrm{sd}}^{\frac{5+p}{4}}\gamma _{\mathrm{w}}^{p-2}r_{\mathrm{s}}(t)^{\frac{1-p}{2}},
\end{eqnarray}
where
$\mathbb{C} =2^{\frac{17-3p}{4}}3^{\frac{p-5}{2}}\pi ^{\frac{3-p}{2}}c^{\frac{-19-3p}{4}}m_{e}^{\frac{-p-3}{2}}q_{e}^{\frac{p+5}{2}}\left( p-2 \right) ^{p-1}\left( p-1 \right) ^{2-p}$, $f_{\rm X}=\left(\nu _{\rm X1}^{1-\beta}-\nu _{\rm X2}^{1-\beta}\right)/\left( \beta -1\right)$, and $\beta =\left( p-1 \right) /2$ (\citealt{Petropoulou2018}).
Obviously, the modulation of X-ray luminosity is completely determined by the variation of the shock distance from the NS. Due to orbital motion and disk precession, interactions between pulsar wind and stellar outflows lead to cyclic changes in $r_{\rm s}(t)$, resulting in the orbital and superorbital modulations of X-ray flux.


\section{Result}

\begin{figure}
  \centering
  \includegraphics[width=0.9\textwidth]{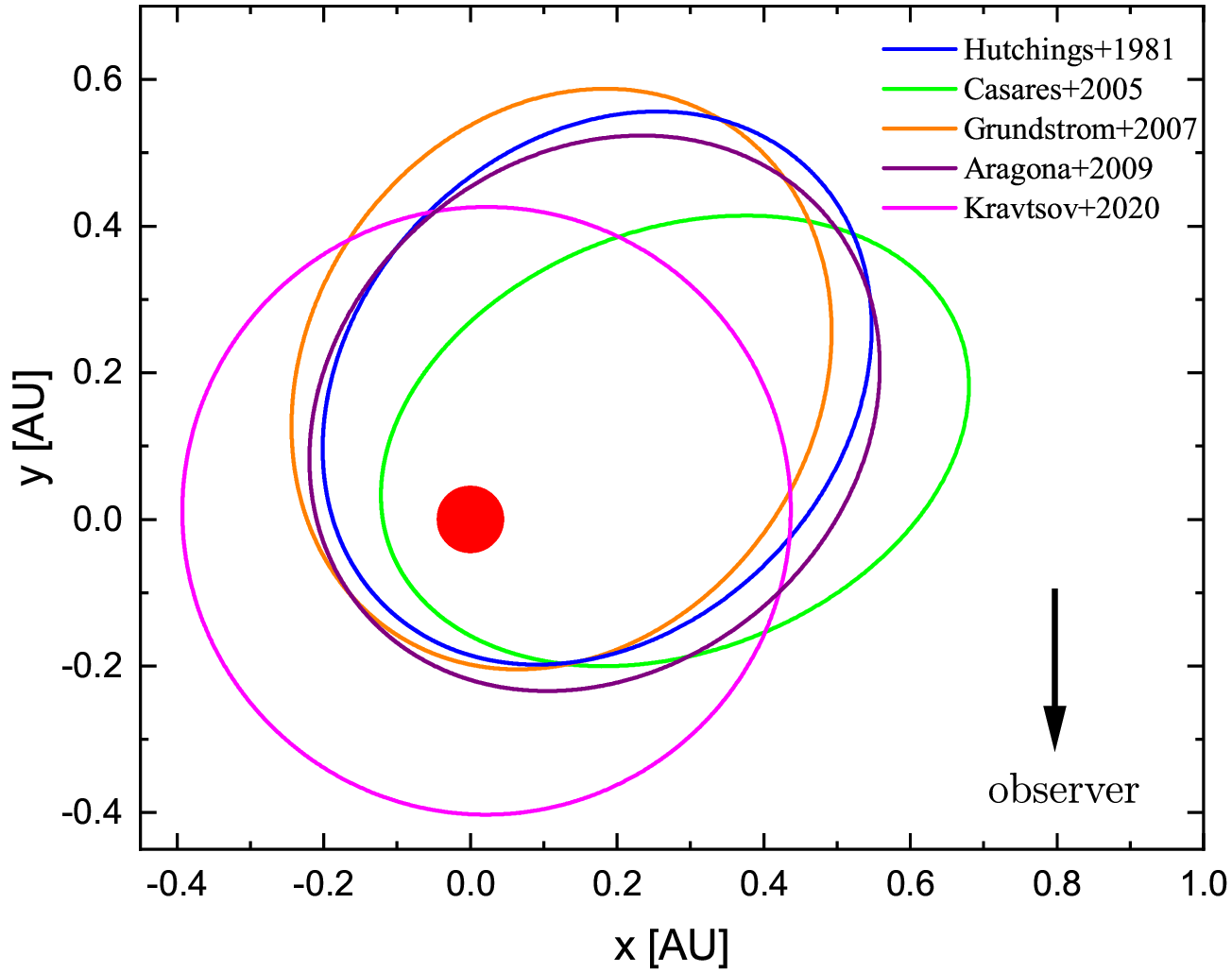}
  \caption{Orbits of \LSI61 with orbital solutions of \citet{Hutchings1981,Casares2005,Grundstrom2007,Aragona2009,Kravtsov2020} as labeled in the plot. The sizes of the star and orbits are to scale.} \label{fig:orbit}
\end{figure}

\begin{figure*}
  \centering
  \includegraphics[width=0.825\textwidth]{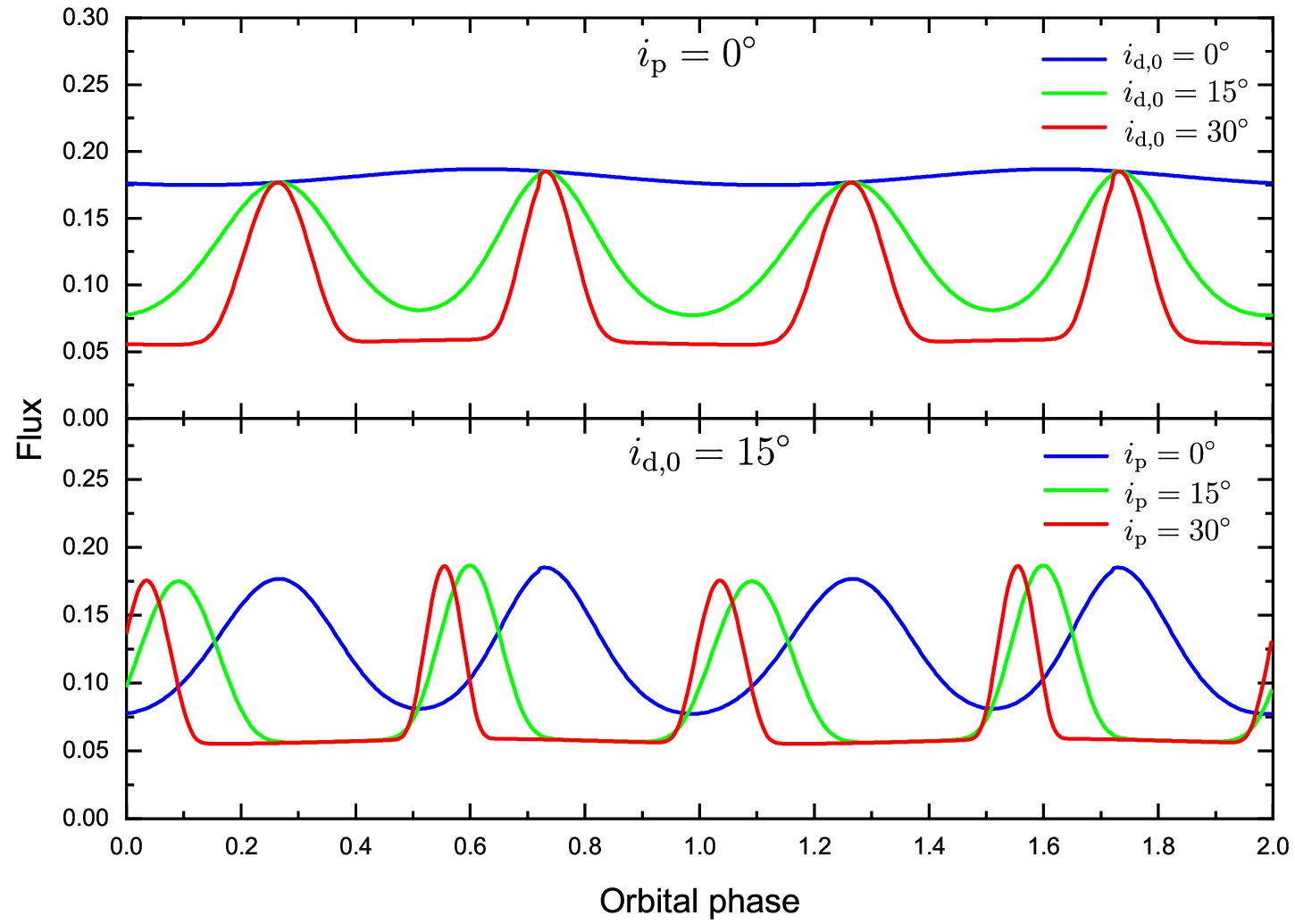}
  \includegraphics[width=0.825\textwidth]{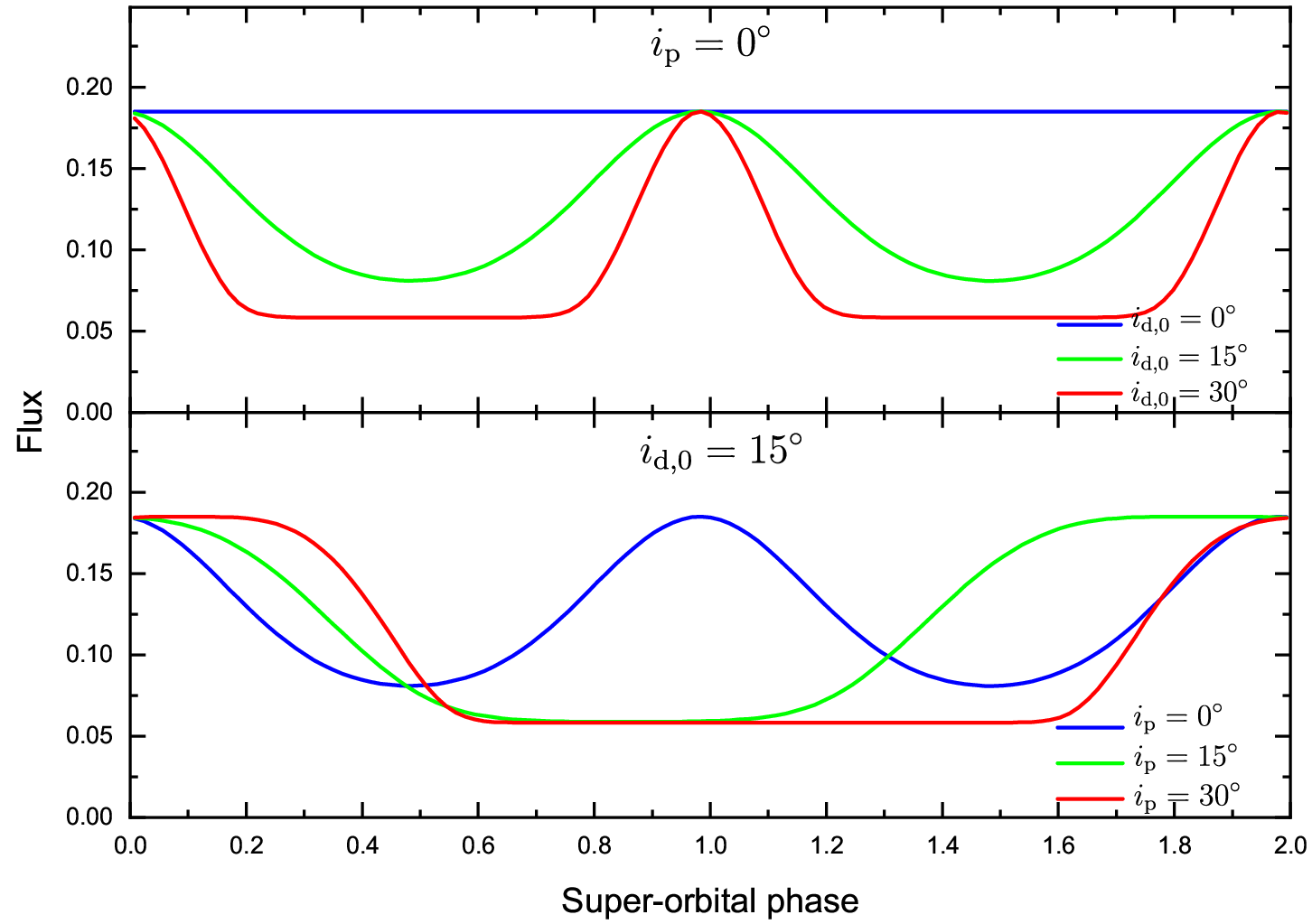}
  \caption{Calculated X-ray flux over the orbital period (top panel) and superorbital period (bottom panel) with different values of $i_{\rm p}$ and $i_{\rm d,0}$ at the phases of $\Theta_{\rm sup}=0.5$ and $\Phi_{\rm orb}=0.5$, respectively. For clarity, we do not include the Doppler boosting effect in this plot.
  } \label{fig:LC_para}
\end{figure*}

\begin{figure*}
  \centering
  \includegraphics[width=0.975\textwidth]{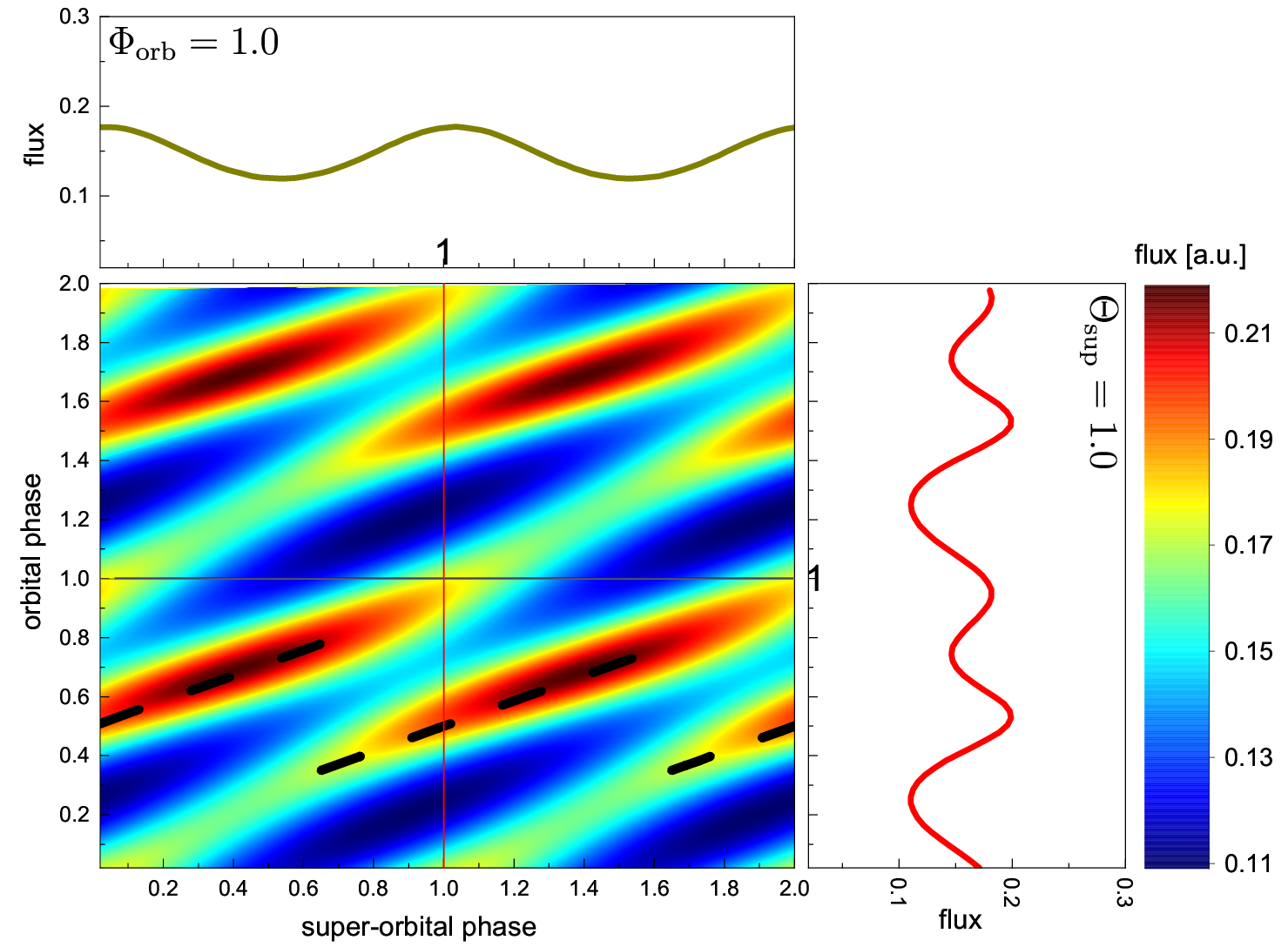}
  \caption{Contour plot of synchrotron X-ray flux as a function of orbital and superorbital phases. The horizontal and vertical histograms show the calculated superorbital light curve at the phase of $\Phi_{\rm orb}=1.0$ and the calculated orbital light curve at the phase of $\Phi_{\rm sup}=1.0$, respectively. The black dashed lines are the observed orbital phase drifting trend of X-ray peak along the superorbital period given by \cite{Chernyakova2012}.} \label{fig:flux_all}
\end{figure*}

\begin{figure*}
  \centering
  \includegraphics[width=0.825\textwidth]{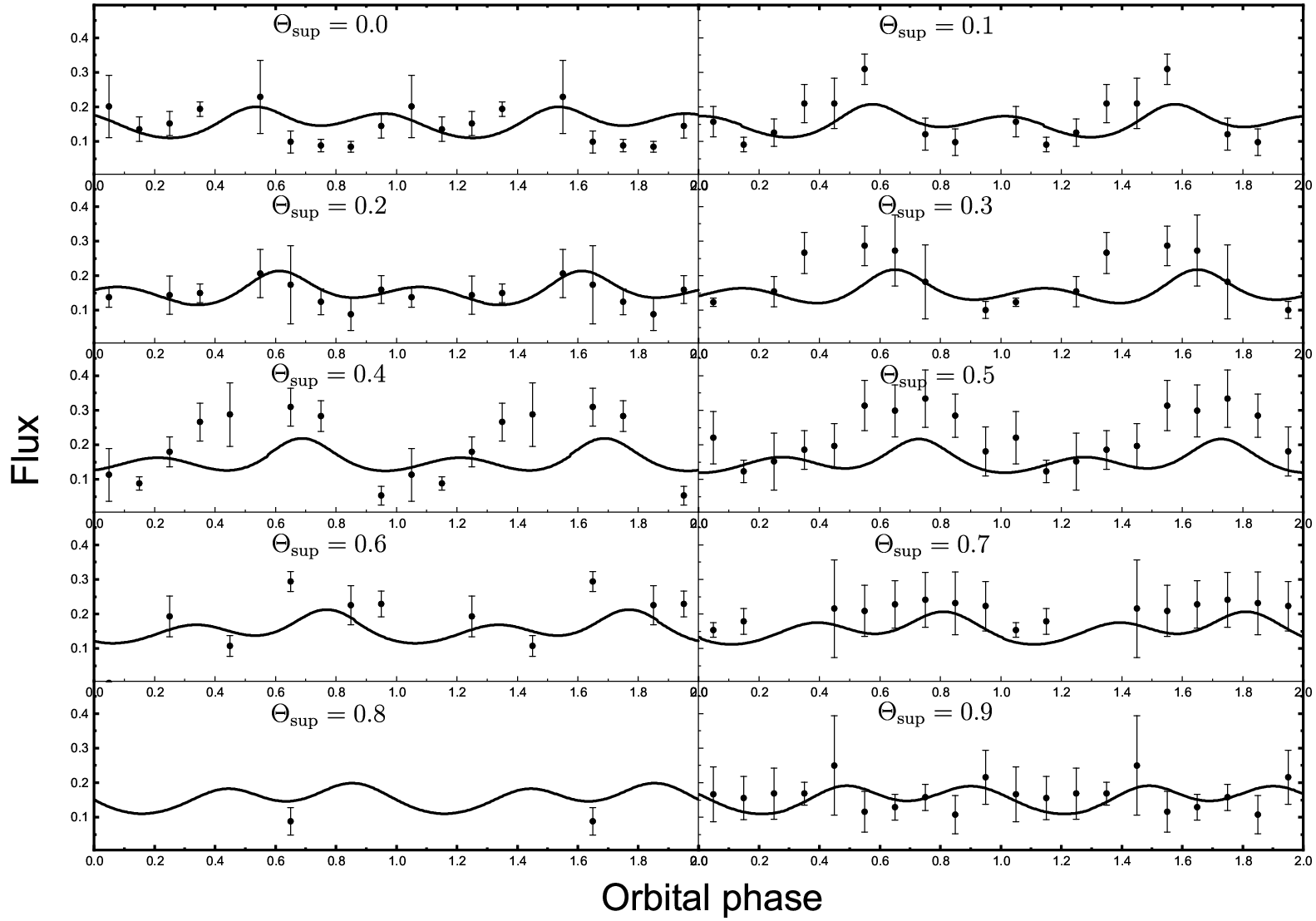}
  \includegraphics[width=0.825\textwidth]{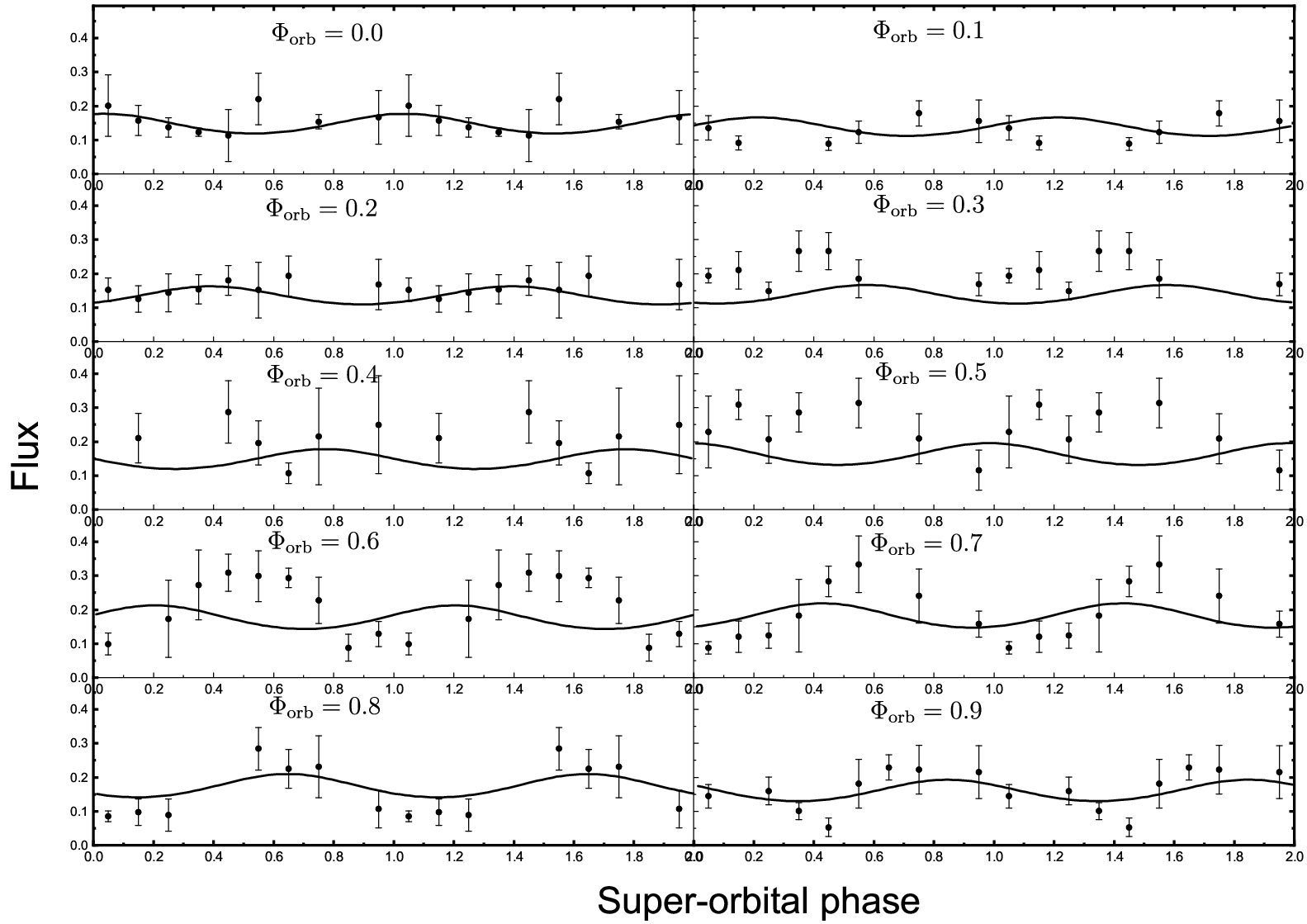}
  \caption{Calculated X-ray flux over orbital period (top panel) and superorbital period (bottom panel) at different phases with comparisons of \textit{Swift}/XRT data taken from \citet{Saha2016}.
  } \label{fig:LC_fit}
\end{figure*}

In this section, we will use the emission model described above to calculate the time-dependent of X-ray flux, and refold it into the orbital and superorbital phases as:
\begin{eqnarray}
    \Phi_{\rm orb} &=& \frac{t-t_0}{P_{\rm orb}}-
    {\rm int}\left(\frac{t-t_0}{P_{\rm orb}}\right),\\
    \Theta_{\rm sup} &=& \frac{t-t_0}{P_{\rm sup}}-
    {\rm int}\left(\frac{t-t_0}{P_{\rm sup}}\right),
\end{eqnarray}
where ${\rm int}(x)$ is the intergerpart of $x$. 
According to Eq.(\ref{LX}), the synchrotron luminosity depends on the following parameters: $\epsilon_{\rm e}, \epsilon_B, L_{\rm sd}, \gamma_{\rm w}, p$ and $r_{\rm s}$. Among them, only $r_{\rm s}$ is time-dependent, which is determined by the orbital solution and the outflow properties of binary stars.
Unfortunately, the orbital elements of \LSI61 are poorly determined, and several orbital solutions are obtained (see Fig. \ref{fig:orbit}).
The observed linear drift of X-ray peak from orbital phase 0.35 to 0.75 in one superorbital period (\citealp{Chernyakova2012}) favors a less elliptical orbit with the periastron phase falling within this range. Therefore, in calculations, we will use the orbital parameters given by \citet{Kravtsov2020}, which specify an eccentricity of $e\simeq0.06$ and a periastron phase of $\Phi_{\rm p}\sim0.6$. These values were derived from fitting the star's optical linear polarization, although they differ significantly from those obtained through measurements of radial velocity variations in spectral lines.
For stellar outflows, we take the typical value of polar wind ram pressure as $p_0=2\times 10^{26}\ \rm{g\cdot cm\cdot s^{-2}}$, the pressure contrast and confinement parameter of the disk as $G=100$ and $m=100$, respectively. The effects of these disk parameters on light curves have been explored in \citet{Chen2019}.
The other model parameters we adopt are as follows:
the NS's luminosity $L_{\rm sd}\simeq5\times10^{36}\ {\rm erg}\cdot {\rm s}^{-1}$, the Lorentz factor of $e^{\pm}$ pairs in pulsar wind $\gamma_{\rm w}=10^4$, and the power-law index of accelerated particles $p=2.1$. The magnetic equipartition parameter and the fraction of spin-down luminosity transferred by shock-accelerated $e^{\pm}$ pairs are $\epsilon_{\rm B}=0.005$ and $\epsilon_{\rm e}=0.1$, respectively. It should be noted that these parameters can be scaled accordingly for other values according to Eq.(\ref{LX}).
The disk precession timescale adopted in calculations is $P_{\rm pre}=2 P_{\rm sup}$ because the emission is expected to be the same when the disk is at the longitude of ascending node $\Omega$ and at $\Omega+180^{\circ}$ (e.g. \citealt{Martin2023}). The observational and model parameters are listed in Table. \ref{tab:para}.

As the NS orbits around the Be companion, the relativistic pulsar wind interacts with the inclined disk twice per orbit. The additional disk pressure pushes the IBS closer to the NS, enhancing the magnetic field and, therefore, producing the asymmetric two-peak profile in the orbital light curve.
Meanwhile, the disk precession leads to cyclical changes in the magnetic field of IBS at each orbital phase, which results in a long-term flux modulation at given orbital phases.
The most uncertain parameters are the inclination angles of the disk normal and the precession axis, which may strongly affect the shapes of light curves. In Fig. \ref{fig:LC_para}, we explore how these angles affect the orbital and superorbital modulations of X-ray flux at the phases of $\Theta_{\rm sup}=0.5$ (top panel) and $\Phi_{\rm orb}=0.5$ (bottom panel), respectively. 
Obviously, in the case of $i_{\rm p}=0^{\circ}$ and $ i_{\rm d,0}=0^{\circ}$, the disk is coplanar with the orbital plane with no precession, then the orbital light curve simply peaks in periastron without long-term modulation at any given orbital phase. With increases of $i_{\rm p}$ or $i_{\rm d,0}$, the asymmetric double-peak features in orbital light curves arise (top panel).
For superorbital light curves (bottom panel), higher inclination angles of the disk normal or the precession axis mean that the NS interacts with the precessing disk only at certain epochs, and the X-ray flux at a given orbital phase remains constant during most of the rest superorbital period. Furthermore, a higher misaligned angle between the precession and orbital axes would also lead to a plateau structure or two sinusoidal peaks in the superorbital light curves.

The X-ray superorbital light curves of \LSI61 show a very smooth sinusoidal-like modulation (e.g., \citealt{Li2012}), suggesting that the Be disk is tilted to the orbital plane with a relatively small angle ($i_{\rm d}\sim10^{\circ}$), and the disk normal vector undergoes precession around the orbital axis ($i_{\rm p}\sim 0^{\circ}$).
This can be explained by the fact that if a disk is misaligned, the CO's tidal torque will tend to align it to the orbital plane (e.g., \citealt{Okazaki2017}).
In Fig.\ref{fig:flux_all}, we show the contour plot of synchrotron-powered X-ray flux as a function of orbital and superorbital phases, with the disk parameters listed in Table. \ref{tab:para}. The horizontal and vertical plots are the corresponding superorbital and orbital light curves at the phases of $\Phi_{\rm orb}=1.0$ and $\Theta_{\rm sup}=1.0$, respectively, as indicated in the main diagram. The black dashed lines are the observed orbital phase drifting trend of the X-ray peak along the superorbital period given by \cite{Chernyakova2012}. 
The corresponding calculated X-ray light curves over orbital and superorbital periods at each phase with comparisons of \textit{Swift}/XRT data are presented in Fig. \ref{fig:LC_fit}. In calculations, we also include the Doppler boosting effect of shocked flows (e.g. \citealp{Dubus2010}) with a very mildly-moving velocity ($v_{\rm sh}\simeq 0.04c$) to give better fittings of observations.
The overall phase drifting trend and calculated light curves are roughly consistent with the observations, although there are still some discrepancies (i.e., the superorbital light curves at the phases of $\Phi_{\rm orb}=0.5-0.7$). The discrepancies could be due to our oversimplifications of the stellar outflow. In particular, the rotation of the disk can induce additional changes of shock geometry and, therefore, to the Doppler boosting flux when the NS is moving around inferior conjunction (i.e., $\Phi_{\rm orb}\sim0.6-0.7$). Furthermore, photoelectric absorption of soft X-rays and inhomogeneities in the stellar outflows can also affect the observed flux (\citealp{Szostek2011,Chernyakova2017}). 

\section{Superorbital periods in Be/X-ray binaries}

\begin{figure*}[b]
  \centering
  \includegraphics[width=0.7\textwidth]{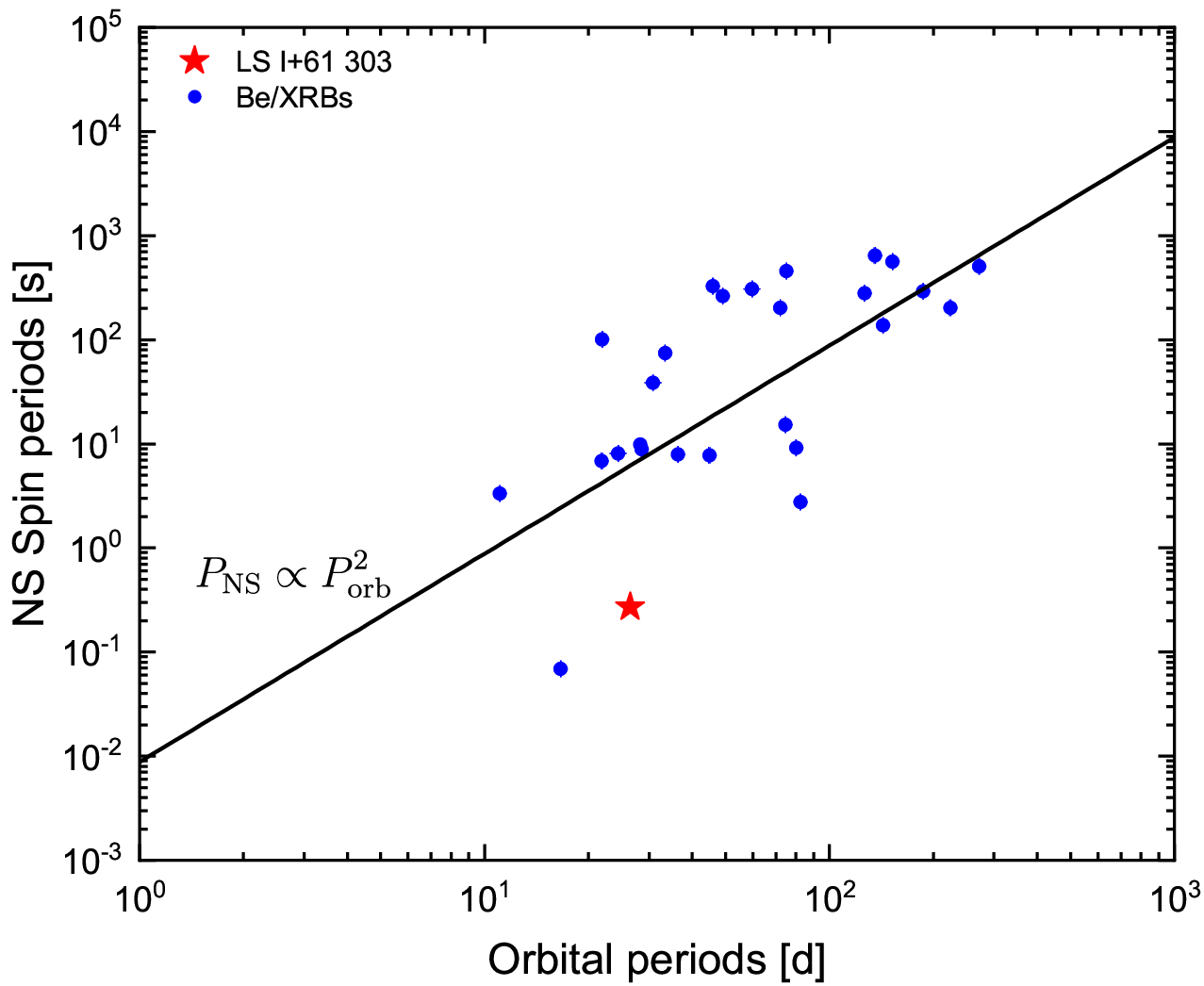}
  \includegraphics[width=0.7\textwidth]{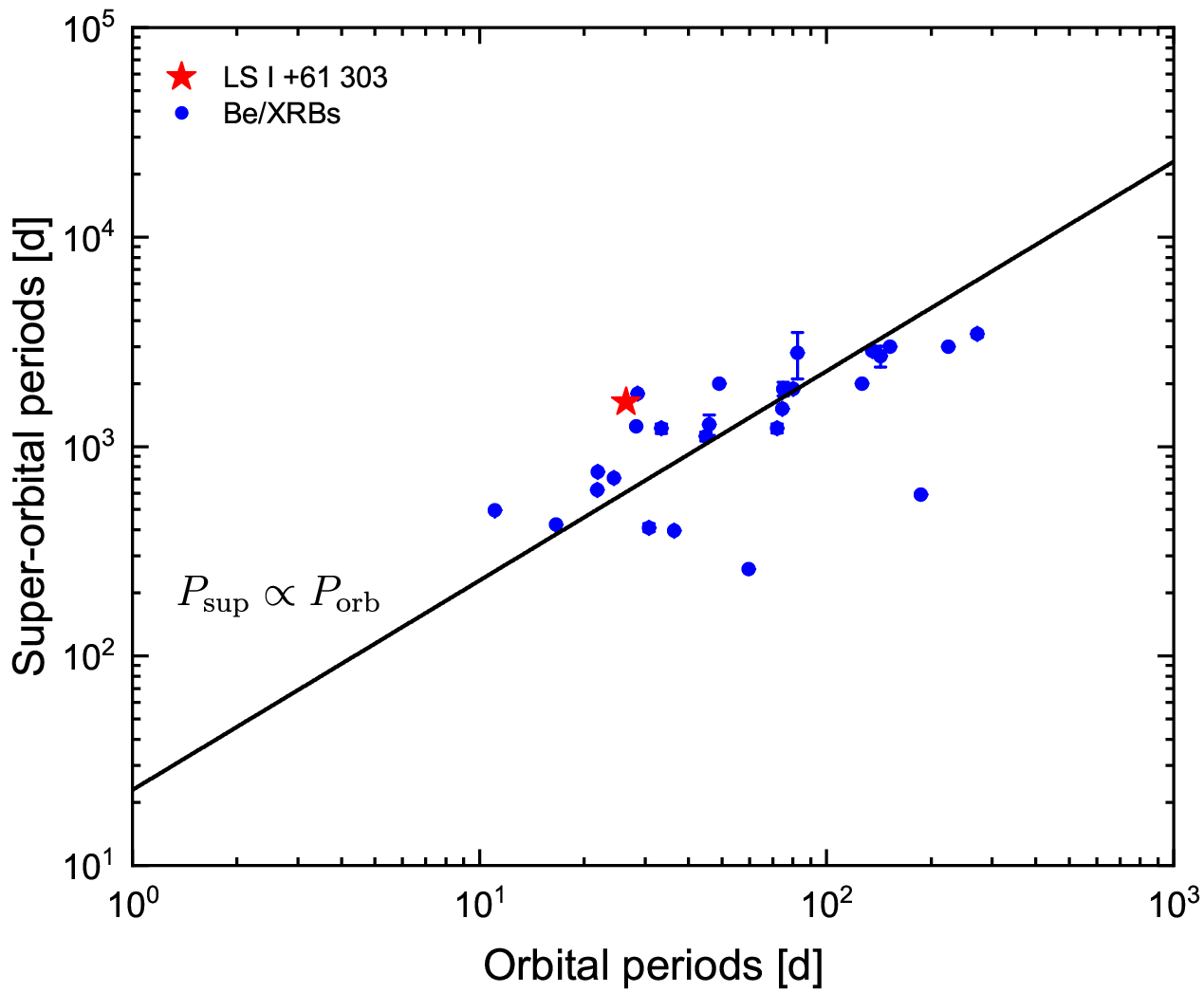}
  \caption{Corbet diagrams (top panel: $P_{\rm NS}-P_{\rm orb}$; bottom panel: $P_{\rm sup}-P_{\rm orb}$) of \LSI61 and other Be/XRBs with superorbital periods as listed in Table. (\ref{tab:Be_XRBs}).
  The black lines in each panel represent the empirical fitting functions with $P_{\rm NS}\propto P_{\rm orb}^2$ and $P_{\rm sup}\propto P_{\rm orb}$, respectively.
  } \label{fig:Be_XRBs}
\end{figure*}

Long-term flux modulations are seen in many types of XRBs.
In Fig. \ref{fig:Be_XRBs}, we plot the Corbet diagrams of \LSI61 and other Be/XRBs with measured values of $P_{\rm NS}, P_{\rm orb}$ and $P_{\rm sup}$, which are listed in Table. \ref{tab:Be_XRBs}.
In the $P_{\rm NS}-P_{\rm orb}$ diagram (top panel), the NS spin-periods of Be/XRBs follow a loose relationship of $P_{\rm NS}\propto P_{\rm orb}^2$ as already noted by \citet{Corbet1986}.
This relationship can be explained by the NS's equilibrium spin-period with a torque of accreted material from the Be disk exerted on the NS (\citealt{Waters1989}):
\begin{eqnarray}\label{Peq}
  P_{\rm{eq}}\simeq
  \frac{2^{3/2}\pi \mu^{6/7}\delta ^{-9/7}}{(2GM_{\mathrm{X}})^{11/7}}\left( \frac{\pi \rho _0}{v_{0}^{3}} \right) ^{-3/7}\left( \frac{GMP_{\mathrm{orb}}^{2} }{4\pi ^2R_{\star}^{2}}\right) ^{(4n-6)/7}
\end{eqnarray}
where $\mu$ is the NS's magnetic moment; $\delta=v_{\rm d,r}/v_{\rm rel}$, $v_{\rm d,r}$ is the disk radial velocity, $v_{\rm rel}$ is the disk velocity relative to the NS's orbital motion; $M=M_{\mathrm{X}}+M_{\star}$ is the total binary mass, $\rho_0$ and $v_0$ are the base density and velocity of the disk at $R_{\star}$, respectively (\citealt{Li1996}). The observed slope in the $P_{\rm NS}-P_{\rm orb}$ diagram around 2 indicates the radial density distribution index of the disk being around $n\simeq3.25$.
For \LSI61, detections of radio pulsations (\citealt{Weng2022}) and lack of accretion signatures in X-rays (\citealt{Sidoli2006}) suggest that the NS is in the rotation-powered state, and the equilibrium relationship may not hold for the system. This may explain why \LSI61 stands as an outlier in the $P_{\rm NS}-P_{\rm orb}$ diagram.

The superorbital periods in Be/XRBs have been suggested to be related to the formation and depletion of Be disks (\citealt{Alcock2001}). However, the linear correlation in the $P_{\rm sup}-P_{\rm orb}$ diagram suggests a binary dynamical origin of the superorbital modulations.
For Be/XRBs, the evolution of the Be disk would be significantly affected by its companion object.
The torque on the disk from the NS companion is (e.g. \citealt{Lai2014}):
\begin{eqnarray}
T_{\rm d} \simeq \frac{GM_{\mathrm{X}}M_{\mathrm{d}}R_{\mathrm{d}}^{2}}{4a^3}\cos i_{\mathrm{d}},
\end{eqnarray}
and the precession rate of the disk normal around the binary axis can be estimated as:
\begin{eqnarray}
\dot{\omega}_{\rm d} \simeq
\frac{1}{4}\frac{M_{\rm X}}{M_{\star}}
\left(\frac{GM_{\star}}{a^3}\right)^{1/2}
\left(\frac{R_{\rm d}}{a}\right)^{3/2}
\cos i_{\mathrm{d}},
\end{eqnarray}
where $M_{\rm d}$ and $R_{\rm d}$ are the mass and size of the disk. Assuming that $R_{\rm d}$ has the same order of the binary separation, and taking the parameters of \LSI61 into the above equation, we can estimate the disk precession timescale with $P_{\rm d}\sim 1000\ {\rm d}$, which agrees with \LSI61 within a factor of $3$.
In addition to the disk precession, the NS companion also exerts a torque on the oblate Be star, resulting in the stellar spin-axis precession (\citealt{Kuhnel2017,Martin2023}). In this case, the disk can also precess like a solid body if its viscosity parameter is high enough. However, this precession timescale is more than $100\ {\rm yr}$ for \LSI61, which can be neglected.



\section{Discussion and Conclusion}
\LSI61 is a unique source for us to investigate dynamical interactions within binary systems and the outflow properties of massive stars.
The most distinctive feature of this binary is its superorbital modulations seen in nearly all wavelengths.
The detection of radio pulsations and the lack of accretion signals strongly favor the wind-interaction scenario instead of the microquasar model.
Assuming synchrotron radiations of the IBS produce X-rays, we calculate the expected flux modulations folded on the orbital and superorbital periods with comparisons of Swift/XRT data.
The asymmetric double-peak profiles, as seen in the orbital light curves of X-rays, are attributed to the interaction between the pulsar wind and the slightly inclined Be disk. Considering the disk precession, the sinusoidal-like long-term modulations in X-rays are reproduced. 
The linear correlation of \LSI61 with other Be/XRBs in the $P_{\rm sup}-P_{\rm orb}$ diagram suggests that the Be disk precession is induced by the tidal effects of the NS companion.

In this paper, we have only focused on the soft X-ray emission properties detected by Swift/XRT, whereas superorbital modulation has been seen in nearly all wavelengths. 
In particular, \citet{Chernyakova2012} performed a long-term multiwavelength analysis (including radio, hard X-ray, and GeV $\gamma$-ray) of \LSI61, and discovered a constant time lag between the radio and X-ray peak. It implies different emitting mechanisms and/or regions in these two bands. It has been suggested that radio emission is likely to be produced by cooled particles at the shock tail at a much larger distance (\citealp{Moldon2013}), and the shock could be deflected due to the orbital motion and disk rotation. Therefore, the radio emission is directly related to the large-scale shock structure, fluid propagation, and cooling processes of particles. Additionally, absorption processes within the binary (including free-free absorption and synchrotron self-absorption) will also affect the observed radio flux (\citealp{Chen2021a}). These factors may explain why the system shows a different modulation pattern compared to X-rays.

\LSI61 also shows energy-dependent variability in the GeV $\gamma$-rays according to the analysis of over 14 years of Fermi/LAT data (\citealt{Chernyakova2023}). In particular, the low-energy (0.1-0.3 GeV) flux is strongly modulated with the orbital period, while modulation in the high-energy (1-10 GeV) is strongly suppressed. It has been suggested that GeV emissions from gamma-ray binaries are likely contributed by several components, including outer-gap emission in pulsar magnetosphere, anisotropic IC scattering within cold pulsar wind and shock emission (e.g., \citealp{Takata2014,Chen2021b}). The emissions from the latter two can be significantly affected by the disk precession. To understand the variability pattern of GeV emission, all of those emission components should be taken into account properly. 

Since its discovery over 45 years ago, \LSI61 has been the subject of intensive observational campaigns across multiple telescopes, resulting in a substantial accumulation of data. This mysterious system exhibits numerous unexpected characteristics that set it apart from other $\gamma$-ray binaries. Despite extensive research efforts, a comprehensive theoretical framework capable of elucidating the multiwavelength flux modulation — spanning from radio to TeV energies — remains elusive. 
Although our current model provides a tentative approach to explain the long-term flux variations in binaries, a more sophisticated emission model should be developed to explain the multiwavelength data.

\begin{acknowledgements}
We are very grateful to the referee for insightful comments on the manuscript and to Dr. Lab Saha for sharing the data of \LSI61. We also thank Prof. Dong Lai and Prof. Weihua Lei for helpful discussions. 
Y.W.Y. is supported by the National Key R\&D Program of China (No. 2021YFA0718500), the National SKA Program of China (No. 2020SKA0120300), the China Manned Spaced Project (No. CMS-CSST-2021-A12) and the National Natural Science Foundation of China (No. 12393811).
J.T. is supported by the National Key Research and Development Program of China (No. 2020YFC2201400) and the National Natural Science Foundation of China (Nos. U1838102, 12173014).
A.M.C. is supported by the China Postdoctoral Science Foundation (No. 2023T160410).
\end{acknowledgements}

\begin{table*}
\caption{Parameters of \LSI61. \label{tab:para}}
\begin{tabular}{l c c c}
\hline
\textbf{Parameter}  &  Symbol  & Value       & Reference\\
\hline
{\it Orbital Parameters} &  &  &\\
Semi-major Axis     &  $a$              & 0.415 \textrm{AU}   &   \\
Eccentricity        &  $e$              & 0.06                & (1) \\
Phase of Periastron & $\Phi_{\rm p}$    & 0.62                & (1) \\
Orbital Period      &$P_{\rm{orb}}$     & 26.496 \rm{d}       & (2) \\
Superorbital Period&$P_{\rm{sup}}$     & 1628 \rm{d}         & (2, 3)   \\

\hline
{\it Stellar Wind and Disk}  &  &  & \\
Ram Pressure of Polar Wind   &$p_0$ &$2\times 10^{26}\ \rm{g\cdot cm\cdot s^{-2}}$  & -\\
Disk-wind Pressure Contrast  &$G$   &$100$  & -\\
Disk Confinement Parameter   &$m$   &$100$  & -\\
Disk Precession Period       &$P_{\rm pre}$ &$2P_{\rm{sup}}$  & -\\
Inclination Angle of Precession Axis        & $i_{\rm p}$    &$0^{\circ}$  & -\\
True anomaly of Precession Axis             & $\phi_{\rm p}$   & arbitrary  & -\\
Inclination Angle of Disk Normal at $t_0$   & $i_{\rm d,0}$  &$10^{\circ}$  & -\\
True Anomaly of Disk Normal at $t_0$        & $\phi_{\rm d,0}$  &45$^{\circ}$  & -\\

\hline
{\it Pulsar Wind and Shock}     &   &  & \\
Pulsar Spin-down Luminosity & $L_{\rm{sd}}$  & $5\times 10^{36}\ {\rm erg\cdot s^{-1}}$   &- \\
Pulsar Wind Lorentz Factor   & $\gamma_{\rm{w}}$  & $ 10^4$  & -\\
Energy Fraction of Accelerated $e^{\pm}$ Pairs  & $\epsilon_{\rm e}$     & $0.1$    & -\\
Equipartition Parameter of Magnetic Field & $\epsilon_{\rm B}$   & $0.005$  & -\\
Distribution Index of $e^{\pm}$ Pairs           & $p$   & $2.1$   & -\\
Velocity of Shocked Flow           & $v_{\rm sh}$   & $0.04c$   & -\\
\hline
\end{tabular}
\\
{{
\textbf{Reference.}
(1) \citealt{Kravtsov2020};
(2) \citealt{Gregory2002a};
(3) \citealt{Massi2016}.
}}
\end{table*}

\begin{table*}
\caption{List of Be/XRBs with superorbital modulations. \label{tab:Be_XRBs}}
\resizebox{\textwidth}{!}{
    \begin{tabular}{l l l l l l l l }
    \hline
        Source  & Stellar Type & $P_{\rm NS} (\rm s)$ & $P_{\rm orb} (\rm d)$ & Error (d) & $P_{\rm sup} (\rm d)$ & Error (d) & References \\
        \hline
        LS I+61 303   & B0V & 0.26915508 & 26.496 & 0.0028 & 1628.0 & 48 & (1-3) \\
        SGR 0755-2933  & B0V & 307.8 & 59.69 & 0.15 & $ \sim$260  & ~ & (4,5) \\
        AX J163904-4642  & BIV-V & 908.79 & 4.23785 & 1.2e-05 & 14.99 & 0.01 & (6,7) \\
        Swift J0243.6+6124  & O9.5V & 9.86 & 28.3 & 0.02 & $\sim$1250 & ~ & (8,9) \\
        H 1145-619  & B1V & 292.4 & 186.68 & 0.05 & $\sim$590 & ~ & (10,11) \\
        IGR J05007-7047  & B2III & 38.551 & 30.776 & 0.005 & 410 & 18 & (12) \\
        RX J0520.5-6932  & O9V & 8.035331 & 24.4302 & 0.0026 & 706.9 & 3 & (13) \\
        A0538-66  & B1III & 0.069 & 16.6409 & 0.0003 & 423.74 & 0.42 & (14) \\
        RX J0059.2-7138  & B1-1.5II-III & 2.76 & 82.37 & 0.07 & 2800 & 700 & (15) \\
        AX J0105-722  & B1-2III-V & 3.34 & 11.07 & 0.01 & 495 & 2 & (15) \\
        XTE J0103-728   & O9.5-B0IV-V & 6.85 & 110.0 & 0.2 & 621 & 4 & (15)  \\
        2S 0050-727   & B1-B1.5IV-V & 7.78 & 44.9 & 0.2 & 1116 & 56 & (15)  \\
        AZV 285     & O9 & 7.92 & 36.41 & 0.02 & 397 & 2 & (15)  \\
        RX J0051.8-7231  & O9.5-B0IV-V & 8.9 & 28.51 & 0.01 & 1786 & 32 & (15)  \\
        RX J0049.2-7311  & B1-B3IV-V & 9.13 & 80.1 & 0.1 & 1886 & 35 & (15)  \\
        RX J0052.1-7319  & O9.5-B0III-V & 15.3 & 74.51 & 0.05 & 1515 & 23 & (15)  \\
        AX J0049-729  & B3V & 74.7 & 33.37 & 0.01 & 1220 & 64 & (15)  \\
        AX J0057.4-7325  & B3-B5Ib-II & 101 & 21.95 & 0.01 & 758 & 6 & (15)  \\
        CXOUJ005323.8-722715  & B1-B2IV-V & 138 & 143.1 & 0.02 & 2700 & 304 & (15)  \\
        1XMMU J005921.0-722317  & B0-B1V & 202 & 71.98 & 0.05 & 1220 & 61 & (15)  \\
        XMMUJ005929.0-723703  & B0-5III  & 202 & 224 & 1 & $\sim$3000 & ~ & (15)  \\
        RX J0047.3-7312  & B1-B1.5V & 264 & 49.06 & 0.02 & $\sim$2000 & ~ & (15) \\
        RX J0057.8-7202  & B0-B2III-V & 280 & 126.4 & 0.2 & $\sim$2000 & ~ & (15)  \\
        XMMU J005252.1-721714.7  & B0-B0.5V & 327 & 45.9 & 0.2 & 1274 & 143 & (15)  \\
        RX J0101.3-7211  & B0.5-B2III-V & 455 & 74.96 & 0.05 & 1886 & 145 & (15)  \\
        AX J0054.8-7244  & B1III-V & 504 & 272 & 1 & 3448 & 119 & (15)  \\
        CXOUJ005736.2-721934  & B0-B2IV-V & 564 & 152.4 & 0.2 & $\sim$3000 & ~ & (15)  \\
        XMMU J005535.2-722906 & B0III & 645 & 135.4 & 0.5 & 2857 & 81 & (15)  \\ \hline
    \end{tabular}
    }
\\{{\textbf{Reference.}
(1) \citealt{Weng2022};
(2) \citealt{Gregory2002a};
(3) \citealt{Massi2016};
(4) \citealt{Doroshenko2021};
(5) \citealt{Richardson2023};
(6) \citealt{Corbet2013};
(7) \citealt{Islam2015};
(8) \citealt{Doroshenko2018};
(9) \citealt{Liu2022} ;
(10) \citealt{White1980};
(11) \citealt{Alfonso-Garzon2017};
(12) \citealt{Vasilopoulos2016};
(13) \citealt{Vasilopoulos2014};
(14) \citealt{Rajoelimanana2017};
(15) \citealt{Rajoelimanana2011}.
}}
\end{table*}

\clearpage



\end{document}